\documentclass[conference]{IEEEtran}
\IEEEoverridecommandlockouts

\usepackage{cite}
\usepackage[T1]{fontenc} 
\usepackage[utf8]{inputenc}
\usepackage{amsmath,amssymb,amsfonts}
\usepackage{algorithmic}
\usepackage{graphicx}
\usepackage{multirow}
\usepackage{textcomp}
\usepackage{xcolor}
\usepackage{amssymb}  
\usepackage{comment}
\usepackage{booktabs} 
\usepackage{array}    
\usepackage{cite}     

\def\BibTeX{{\rm B\kern-.05em{\sc i\kern-.025em b}\kern-.08em
    T\kern-.1667em\lower.7ex\hbox{E}\kern-.125emX}}
\begin{document}

\title{PulseFi: A Low Cost Robust Machine Learning System for Accurate Cardiopulmonary and Apnea Monitoring Using Channel State Information\\









}


\author{
    \IEEEauthorblockN{Pranay Kocheta\textsuperscript{1}, Nayan Sanjay Bhatia\textsuperscript{2}, Katia Obraczka\textsuperscript{2}} 
    \IEEEauthorblockA{\textsuperscript{1}Independent, Boston, Massachusetts, USA \\ pranayko021@gmail.com} 
    \IEEEauthorblockA{\textsuperscript{2}University of California, Santa Cruz, Santa Cruz, California, USA \\ \{nbhatia3, obraczka\}@ucsc.edu}
}

\maketitle

\begin{abstract}
Non-intrusive monitoring of vital signs has become increasingly important in a variety of healthcare settings. In this paper, we present PulseFi, a novel low-cost non-intrusive system that uses Wi-Fi sensing and artificial intelligence to accurately and continuously monitor heart rate and breathing rate, as well as detect apnea events. PulseFi operates using low-cost commodity devices, making it more accessible and cost-effective. It uses a signal processing pipeline to process Wi-Fi telemetry data, specifically Channel State Information (CSI), that is fed into a custom low-compute Long Short-Term Memory (LSTM) neural network model. We evaluate PulseFi using two datasets: one that we collected locally using ESP32 devices and another that contains recordings of 118 participants collected using the Raspberry Pi 4B, making the latter the most comprehensive data set of its kind. Our results show that PulseFi can effectively estimate heart rate and breathing rate in a seemless non-intrusive way with comparable or better accuracy than multiple antenna systems that can be expensive and less accessible.  

\end{abstract}

\section{Introduction}
Non-intrusive monitoring of vital signs (such as heart, breathing rate, and sleep apnea) has become increasingly important, particularly for home care, elderly care, and managing chronic conditions. As the global population ages and chronic disease rates increase, there is a growing need for continuous and accurate vital sign monitoring systems that can be easily deployed across the healthcare continuum, including hospitals, long-term care and home care settings~\cite{chen2023digital}. Breathing and heart rate provides critical information about an individual's respiratory and cardiovascular health. Furthermore, detection of apnea, characterized by temporary pauses in breathing (typically lasting 10 seconds or longer)\cite{AASMD_2015_SleepApnea}, is critical as conditions like sleep apnea affect millions worldwide and can lead to serious health complications if undiagnosed \cite{punj2017sleep}. Thus, non invasive monitoring of these cardiopulmonary variables is necessary. 

Traditional approaches for vital sign monitoring have relied heavily on contact-based sensors such as pulse oximeters, heart rate belts, chest straps, or highly specialized medical equipment, such as polysomnography (PSG) or electrocardiogram (ECG) devices. However, these methods can be intrusive, uncomfortable, and impractical for long-term use, affecting sleep quality and natural behavior~\cite{galli2022overview, huang2023challenges, sartor2018methodological}. To address this, a number of approaches have proposed non-contact vital monitoring methods, specifically computer vision-based systems. However, these visual systems raise significant privacy concerns and often do not function in low light conditions or when the subject is obstructed~\cite{huang2023challenges, selvaraju2022continuous, kumar2015distance}.

In response to these limitations, researchers have turned to wireless sensing as a promising alternative to non-invasively monitoring heart and respiration rates~\cite{soto2022survey, khamis2018cardiofi, solgun2023real, wang2020csi, tsubota2021biometric, liu2022human_jbhi}. Wi-Fi signals can penetrate walls and furniture, are ubiquitous in most indoor environments, and do not raise the same privacy issues as camera-based systems. Specifically, by analyzing the Channel State Information (CSI) of Wi-Fi signals, which captures how the signal propagates between transmitter and receiver, it is possible to detect small chest movements associated with both breathing cycles and heartbeats~\cite{ma2019wifi_survey, soto2022survey, zhang2023wital}, and even the stoppage of breathing during apnea \cite{liu2014wisleep, adib2015smart}.

Wi-Fi infrastructure for vital sign monitoring should ideally satisfy the following requirements:
\begin{itemize}
\item \textbf{Robustness to Real-World Environments:}
The amplitude or phase of CSI data can become extremely weak and noisy when transmitters and receivers are farther apart, leading to high error~\cite{ma2019wifi_survey, soto2022survey, wang2020csi}. Respiratory signals are also considerably stronger than cardiac signals, often masking subtle heartbeat variations. Apnea detection requires robust differentiation between normal breathing, shallow breathing, and the absence of breathing. Additionally, signal processing techniques are generally based on peak detection methods\cite{soto2022survey}. While these approaches can be effective under ideal conditions, they often struggle to handle complex, noisy environments encountered in the real-world.
 
\item \textbf{Reliability:}
Medical systems must be reliable and thus require thorough testing and evaluation under diverse conditions. A large number of participants and different scenarios are required for reliability and real-world applicability.

\item \textbf{Accessibility:} Low computational requirements are necessary to ensure ubiquitous adoption~\cite{compton2018access, ansermino2013universal} of the system, along with accuracy. By minimizing the system's computational complexity, it becomes more feasible to deploy in low-cost, everyday devices, making it accessible to a larger population. 

\item \textbf{Accessibility:}
Extracting CSI from Wi-Fi chipsets~\cite{gringoli2021axcsi} can be challenging due to hardware and firmware limitations as well as lack of support for legacy systems. This may create obstacles for adoption. 
Hence, open-source systems like Nexmon CSI and ESP-32~\cite{Hern2006:Lightweight,gringoli2021axcsi} provide low cost and ubiquitous solution to make CSI-based applications available to broader audiences. 

\item \textbf{Accuracy:} The system must provide clinically relevant accuracy to be a viable alternative to traditional methods. 
\end{itemize}


To the best of our knowledge, 
PulseFi is the first system to address all these requirements for integrated cardiopulmonary and apnea monitoring. It uses low-cost commodity devices, making it more accessible and cost-effective. It includes a signal processing pipeline to process CSI data which is then fed into a custom low-compute Long Short-Term Memory (LSTM) neural network model. In light of our previous work~\cite{11096342}, where we first introduced an early iteration of PulseFi, the contributions of this paper can be summarized as follows:
\begin{itemize}
    \item We describe PulseFi's current design and architecture in detail, including its unified CSI data processing pipeline that can be used for heart rate and breathing rate monitoring, as well as apnea event detection, as well as other applications.  
    \item We demonstrate that PulseFi can achieve high‑accuracy vital‑sign monitoring without exploiting phase information. By relying solely on the absolute amplitude of the received signal, our method works with inexpensive single-antenna hardware and obviates the need for complex phase difference measurements, yet still delivers reliable accuracy.
    \item We also zoom in on PulseFi's compact LSTM models, whose low computing requirements allow them to be deployed on resource-constrained devices like the ESP32 and the Raspberry Pi, enabling accessible, low-cost real-time, continuous monitoring.
    
    \item We evaluate PulseFi using two CSI datasets: (1) one of them was collected using ESP32 devices with seven participants in a semi-controlled indoor environment across varying distances, (2) the other CSI dataset contains data from 118 participants in 17 positions/activities, making it the most comprehensive dataset of its kind. 
\end{itemize}

The remainder of this paper is organized as follows: In Section II, we highlight related work, setting the stage for our contributions. Section III provides an overview of PulseFi, while Sections IV, V, and VI describe PulseFi's functional components in detail. Section VII outlines the experimental methodology we use to evaluate PulseFi and Section VIII presents the results of our experiments - summarizing our findings and highlighting key insights. Finally, Section IX concludes the paper with some directions of future work.


\section{Related Work}
Current methods for monitoring heart, breathing rate, and apnea can be classified into two main categories: contact based and non-contact based approaches. We use these categories to describe existing work related to PulseFi.

\subsection{Contact-Based Sensing}
Clinical standards like ECG, PPG, chest belts (for respiration), and PSG offer high accuracy for heart rate, breathing patterns, and apnea diagnosis \cite{kapur2017guideline}. However, they often involve multiple sensors, can be costly, and are uncomfortable for long-term or home monitoring \cite{galli2022overview, sartor2018methodological}. Systems like chest straps, pulse oximeters, and smartwatches, while more accessible, can still cause discomfort over extended periods and may not provide comprehensive respiratory data for apnea assessment \cite{massaroni2019wearable}.

\subsection{Non-Contact Sensing}
Non-contact methods for monitoring heart and breathing rates have gained significant attention due to their potential for non-intrusive and continuous health monitoring. Such techniques eliminate the discomfort and limitations that come with contact-based sensors. Common non-contact approaches include camera-based monitoring and wireless signals such as Wi-Fi.

\subsubsection{Camera}
Camera-based techniques, such as remote photoplethysmography (rPPG), analyze subtle changes in skin color to estimate heart and respiratory rates. While offering a non-invasive monitoring option, these methods face challenges including high costs associated with advanced imaging equipment. Thermal imaging cameras can also be used to detect subtle temperature variations associated with blood flow, which allows for heart rate, breathing rate, and apnea monitoring, though they are typically more expensive than traditional cameras. They also raise privacy concerns as they rely on continuous video recording. In addition, factors such as lighting conditions and skin tone variations can affect measurement accuracy \cite{huang2023challenges, manullang2021thermal, vats2022early}.

\subsubsection{Radar}
Radar technologies, including Doppler, Ultra-Wideband (UWB), and Frequency-Modulated Continuous Wave (FMCW) radars, detect minute chest movements caused by cardiac and respiratory activity. While they have proven to be effective and are able to achieve high levels of accuracy, these systems often require complex, custom setups that come with high costs, making them impractical for various deployments ~\cite{ishrak2023doppler, islam2022radar}. Their sensitivity to environmental factors and need for specialized equipment further limit their widespread adoption outside high-end clinical settings \cite{raheel2024contactless_survey}. 


\subsubsection{Wireless Signals}
The two most common telemetry metrics in wireless signals are Received Signal Strength Indicator (RSSI) and Channel State Information (CSI).
RSSI based methods use variations in radio signal strength to detect physiological movements associated with respiration and heartbeat. However, these methods are highly sensitive to environmental noise and interference, which can significantly degrade accuracy \cite{khalili2020wi_sensing, ma2019wifi_survey}. Extreme motion sensitivity and inadequate spatial resolution make RSSI based systems less reliable for precise vital sign monitoring or apnea detection. 

CSI captures the environmental conditions by providing the phase and amplitude of different subcarriers. This enables the system to be reliable against scattering, fading, and power delays\cite{ma2019wifi_survey}. 
\begin{itemize}
    \item \textbf{Breathing Rate (BR) Estimation}: Early CSI work focused on BR estimation using bandpass filtering and spectral analysis (e.g., FFT) to find dominant frequencies \cite{liu2015wifiu}. More recent works use machine learning for improved accuracy \cite{zhang2018finegrained_resp}.
    \item \textbf{Heart Rate (HR) Estimation}: While CSI-based hear rate estimation follows principles similar to that of breathing rate, it is more challenging due to the weaker cardiac signal compared to respiration~\cite{soto2022survey}.
    \item \textbf{Apnea Detection}: CSI has also been explored for apnea detection by identifying the absence or significant reduction of respiratory movements ~\cite{ liu2014tracking_sleep}, with some systems using machine learning on CSI features to classify apnea events~\cite{wang2018wispiro}.
\end{itemize}

Despite potential, current CSI systems have limitations. Many rely on specialized, now-discontinued hardware (e.g., Intel 5300 NIC) with multi-antenna setups to capture phase-difference information~\cite{khamis2018cardiofi, wang2020csi, liu2022human_jbhi}. Evaluations are often on a single individual or a small, homogeneous group. Other limitations include proprietary hardware and no variation in subject posture/position or amount of time over which data is collected (temporal duration)~\cite{sun2024noncontact_hr, khamis2018cardiofi, tsubota2021biometric, wang2020csi, liu2022human_jbhi, solgun2023real}. Machine learning, while powerful, has seen limited exploration for HR monitoring~\cite{soto2022survey}, and robust, computationally light models for combined cardiopulmonary and apnea monitoring on commodity hardware are still an open area. The work in ~\cite{gouveia2024parameter_tuning} uses the EHealth CSI dataset (one of our evaluation datasets) for HR monitoring, but shows inconsistencies between postures and activities. Our work aims to improves upon the current state-of-the-art by also including BR and apnea detection, and demonstrating more consistent performance.

\section{PulseFi: System Overview}
\begin{figure*}[!htbp]
    \centering
    \includegraphics[width=0.7\linewidth]{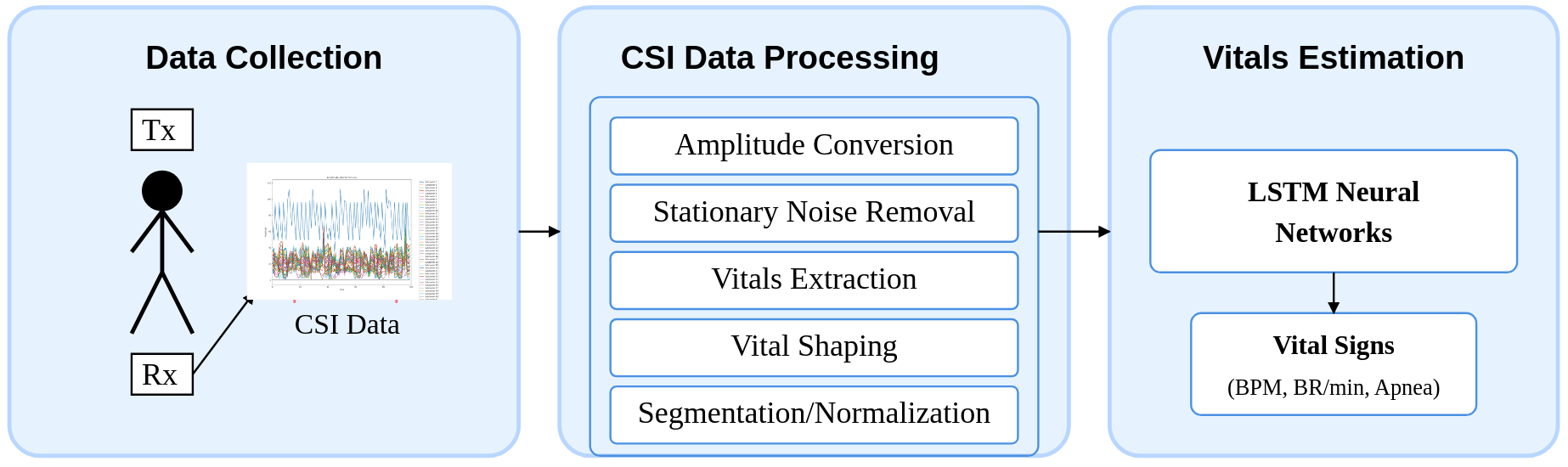}
    \caption{PulseFi System Architecture}
    \label{fig:system_arch}
\end{figure*}
PulseFi's goal is to provide a low-cost, accessible end-to-end solution to non-intrusive, continuous vital sign monitoring. Its design prioritizes accessibility and real-world usability while maintaining high accuracy. It leverages the widespread availability of Wi-Fi deployments and uses Wi-Fi telemetry to perform non-intrusive vital sign monitoring. Fig \ref{fig:system_arch} illustrates PulseFi's system architecture which consists of three main components: Wi-Fi telemetry collection, specifically Channel State Information (CSI) using commodity Wi-Fi devices, a CSI signal processing pipeline, and a custom lightweight Long Short-Term Memory (LSTM) neural network for vital sign estimation. 

\subsection{CSI Data Collection}
CSI data is collected using commodity Wi-Fi devices, either an ESP32 or Raspberry Pi 4, configured as transmitter and receiver pairs. The transmitter continuously broadcasts Wi-Fi packets while the receiver extracts CSI measurements from the received packets. When a subject is placed between the devices, their cardiopulmonary activity affects the wireless channel through minuscule chest movements. These modulations are shown as temporal variationxs in the CSI amplitude across subcarriers explained in the processing pipeline.

\subsection{CSI Processing Pipeline}
PulseFi's CSI processing pipeline was designed to isolate and extract the subtle CSI variations caused by heartbeats and breathing while removing environmental noise and interference. The pipeline consists of five stages:
\begin{itemize}

\item The signal amplitude is extracted from the raw CSI data to capture amplitude variations.
\item The stationary (DC) component of the amplitude is removed to center the signal around zero, allowing us to isolate only the meaningful amplitude variations. 
\item Frequencies corresponding to typical heart rates or breathing rates (0.8 to 2.17 Hz, corresponding to 48 to 130 beats per minute (BPM); 0.1 to 0.5 Hz, corresponding to 6 to 30 breaths per minute (BRPM), with adaptability for the detection of apnea as described in Section ~\ref{sec:csi_processing})  are extracted using a bandpass filter.
\item A Savitzky-Golay filter is used for final smoothing and to reduce leftover noise. Finally, the resulting data are segmented into overlapping windows, and each window is normalized before being fed to PulseFi's LSTM module. 
\end{itemize}

\subsection{LSTM Network}
\label{sec:LSTM}
The heart rate estimation uses a Long Short-Term Memory (LSTM) neural network. We chose to use an LSTM model for four key reasons. LSTM models specialize in processing sequential and temporal data\cite{shang2021lstm_cnn}, which is the case of CSI variations captured for the purposes of measuring heart and breathing rate. Second, the model requires minimal computational resources and hardware making it a low-cost, more accessible system \cite{krishna2018lstm_activity}. Third, because it uses a deep learning based approach, LSTM models exhibit higher immunity to noisy data \cite{laitala2020robust_ecg}. Finally, unlike traditional convolutional neural networks (CNN), LSTMs can naturally handle variable length input sequences allowing the system to adapt to different monitoring durations and requirements \cite{ma2019wifi_survey}.

\section{CSI Data Processing} \label{sec:csi_processing}

As shown in Fig~\ref{fig:system_arch}, PulseFi's CSI data processing component consists of several steps to improve CSI signal quality and extract features for vital estimation and apnea detection. These steps aim at isolating the subtle variations in the CSI signal that correspond to changes caused by the heart, chest movements, or cessation of breathing. 

\subsubsection{Amplitude Conversion}

\begin{figure}
    \centering
    \includegraphics[width=\columnwidth]{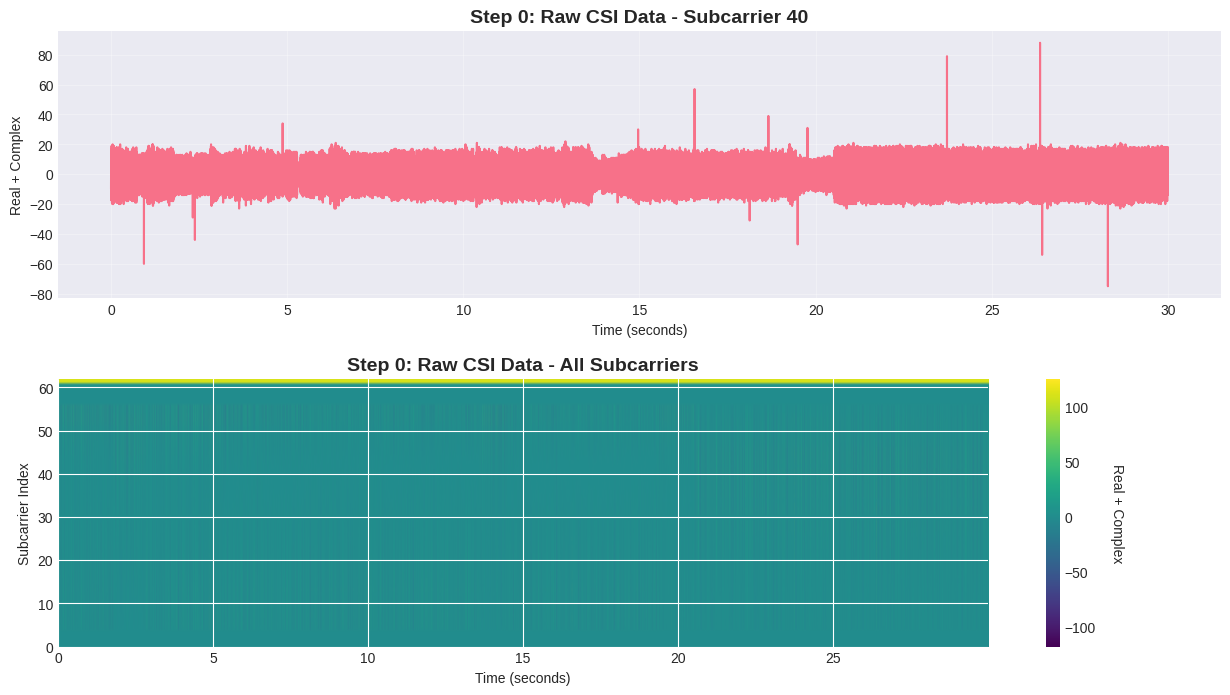}
    \caption{Raw CSI data}
    \label{fig:raw_csi}
\end{figure}

Raw CSI data (Fig~\ref{fig:raw_csi}) is commonly represented as complex numbers that contain both amplitude and phase information as described in Equation~\ref{eq:amplitude}. We extract the amplitude portion (Fig 3) of the CSI using Equation~\ref{eq:amplitude} since variations in amplitude of the signal are directly related to physical movements caused by heartbeats and chest expansion. 

\begin{equation}
A = \sqrt{(\text{Re(CSI)})^2 + (\text{Im(CSI)})^2}
\label{eq:amplitude}
\end{equation}
where $A$ represents the amplitude of the signal, while $\text{Re(CSI)}$ and $\text{Im(CSI)}$ are the real and imaginary parts of the CSI data, respectively.

Because ESP32 and Raspberry Pi use a single antenna, any absolute phase measurement is dominated by noise. In contrast, systems with multiple antennas can take advantage of phase-difference information, which is far more informative. In our work, we deliberately avoid phase altogether. Instead, we rely solely on the absolute amplitude of the received signal, demonstrating that high‑accuracy vital‑sign monitoring is achievable even with inexpensive single-antenna hardware. Our key contribution is therefore to show that, without the need for sophisticated phase-difference measurements, PulseFi can still deliver reliable accuracy.

\begin{figure}
    \centering
    \includegraphics[width=\columnwidth]{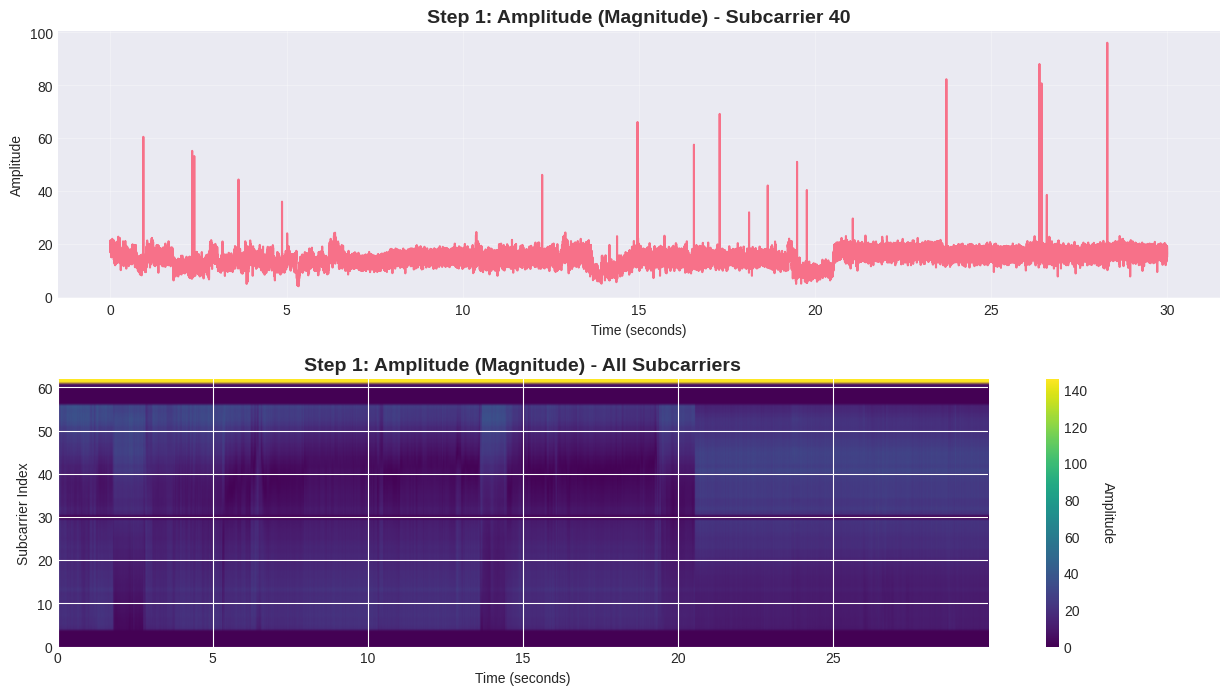}
    \caption{Signal amplitude extracted from raw CSI data}
    \label{fig:amp_csi}
\end{figure}

\subsubsection{Stationary Noise Removal}
Raw CSI data typically include different forms of noise caused by hardware imperfections, noise from the communication channel, etc. 
To remove these static interferences, we remove the signal's DC (direct current) component.~\cite{wang2014wecanhear}. This helps isolate the dynamic parts of the signal as in the human and their vitals. 

\subsubsection{Pulse/Respiration Extraction}
One key challenge in measuring vitals from CSI data is to isolate the rhythmic movement of the heart and chest from various other biological and environmental factors. To address that, we use the Butterworth bandpass filter~\cite{jagtap2012butterworth}. For heart rate, we listen to frequencies between 0.8 Hz and 2.17 Hz (48-130 bpm). For breathing rate, PulseFi captures frequencies between 0.1 Hz and 0.5 Hz (6-30 breaths/min). In the case of apnea detection, the lower frequency band is modified to 0 Hz to show the absence of signal within this typical breathing range, or specific patterns indicating cessation. The filter also has no ripples in the passband or stopband, which prevents artificial oscillations in the filtered signal. In our implementation, we empirically determined that a third-order bandpass filter (N=3) gives sufficient steepness in the transition between passed and blocked frequencies while still being computationally light.

\subsubsection{Pulse/Respiration Shaping}

To further improve the signal and reduce high-frequency noise while preserving important features, we applied a Savitzky-Golay filter~\cite{schafer2011savitzky} as illustrated in Fig 4. This filter performs local polynomial regression on the CSI values defined in Equation~\ref{eq:savgol}, providing smoothing that is very effective at maintaining the shape of physiological signals. In our current implementation, we use a window length of 15 samples and a polynomial order of 3, which were empirically derived to try to balance noise reduction with signal preservation. 

\begin{equation}
y_{\text{smoothed}}[i] = \sum_{k=-m}^{m} c_k \cdot y[i+k]
\label{eq:savgol}
\end{equation}
where $c_k$ are the Savitzky-Golay filter coefficients, $y$ is the input signal, $y_{\text{smoothed}}$ is the smoothed output, i denotes the current sample index, and
k is the relative offset within the smoothing window.

\begin{figure}
    \centering
    \includegraphics[width=\columnwidth]{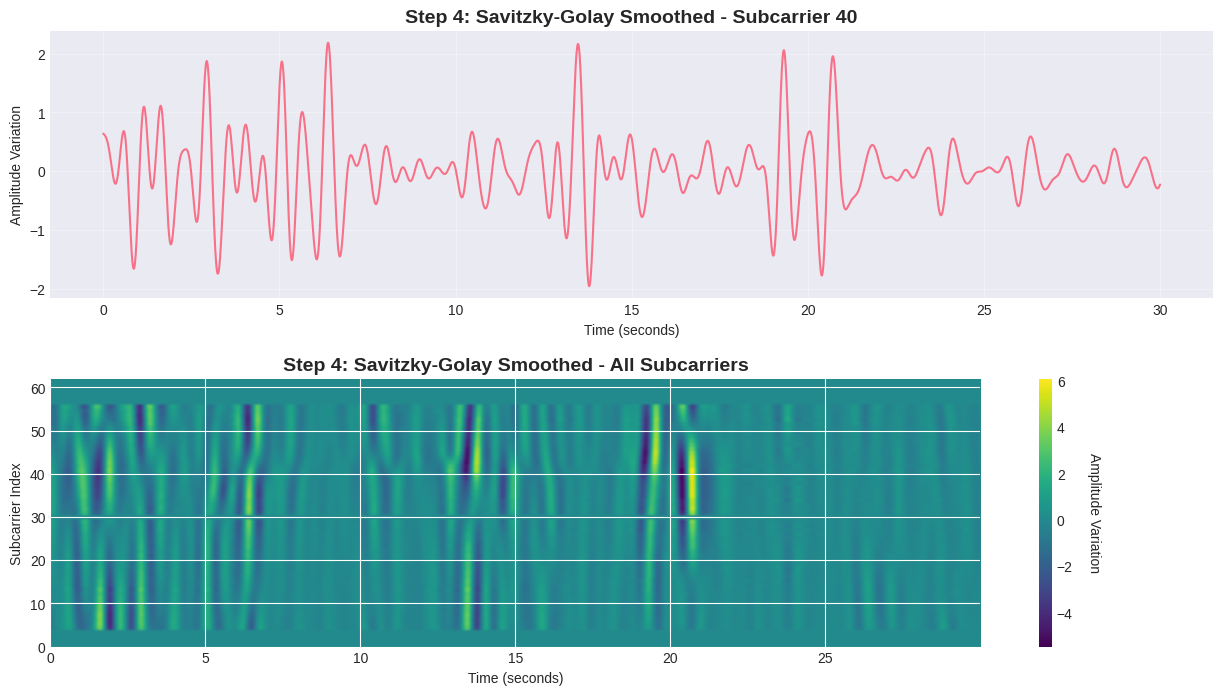}
    \caption{CSI data after filtering for HR}
    \label{fig:csi_post_filter}
\end{figure}

\subsubsection{Data Segmentation and Normalization}
To prepare the data for input into PulseFi's LSTM heart rate estimator, we segment the CSI into overlapping windows, where each window contains a fixed number of consecutive packets. Consecutive windows overlap, meaning that for a window length of 100 packets, the first window will be packets 1-100 and the second 2-101. The temporal duration of each window is calculated as the window size in seconds multiplied by the sampling rate. For example, with the ESP-CSI dataset which has a sampling rate of 80 Hz, a 5s window size would have length 400 packets. This segmentation allows the model to capture the temporal dependencies in the signal. Each segment is then normalized using standardization, which helps to achieve faster convergence during model training and ensures that all features are on a similar scale. 
\begin{equation}
A_{\text{normalized}} = \frac{A - \mu_A}{\sigma_A}
\label{eq:normalization}
\end{equation}
where $A_{\text{normalized}}$ is the resulting normalized amplitude and $\mu_A$ represents the mean amplitude across all time samples for each subcarrier.

 \section{Vital Sign Estimation}

Processed CSI data is fed to PulseFi's vital sign estimation module which uses a Long Short-Term Memory (LSTM) neural network to continuously estimate heart rates or breathing rates. As discussed in Section~\ref{sec:LSTM}, LSTM-based models are well-suited for this task as they can find long-term dependencies in time series data~\cite{shang2021lstm_cnn, krishna2018lstm_activity, laitala2020robust_ecg, ma2019wifi_survey}. We describe PulseFi's LSTM architecture in detail below as well as how it is trained to estimate heart rate, breathing rate and detect apnea.

\subsubsection{PulseFi's LSTM Architecture}
LSTMs are able to extract the long- and short-term features of datasets using {\it memory cells} and {\it gates}. Memory cells store long-term and short-term dependencies of sequential data. For instance, running will cause sudden spikes in vitals that can be handled by short-term memory, and information such as resting values can be preserved by long-term memory cells. Gates are responsible for retaining and discarding CSI data features. There are three types of gates: forget gates discard irrelevant information, input gates decide what information is retained and output gates represent the long- and short-term features. 

Fig 5 shows PulseFi's LSTM-based vital sign estimation architecture, which is designed to effectively handle the sequential time nature of CSI data. The number of layers for each block is empirically chosen to strike a balance between accuracy and real-time prediction. Below, we describe each functional component in PulseFi's LSTM architecture:

\begin{itemize}
    \item CSI Ingestion: Receives processed CSI Data in chunks from PulseFi's CSI processing module based on the selected window size.
    \item Pattern Discovery (64 LSTM Layers): Finds initial patterns to correlate CSI derived amplitude with vital signs.
    \item Feature refinement (32 LSTM Layers): Finds the most important features of CSI data for vital estimation.
    \item Dimension Reduction (16 Dense Layers): Provides the selected features to estimate the desired vital sign and drops any other redundant or irrelevant information. 
    \item Overfitting Prevention: Generalizes the model for different scenarios. 
    It uses a "dropout" mechanism which deactivates a percentage of neurons during training. The goal of this is to reduce overfitting by forcing the network to learn more robust features. As shown in Fig 5, we incorporate this block after every LSTM block. In our current LSTM implementation, we empirically set the dropout rate to $0.2$. 
    \item Non-Linear Pattern Recognition: Since patterns are rarely linear in nature, we use a non-linear activation function (ReLU) that helps the LSTM model to find intricate patterns by introducing non-linearity. ReLU outputs the input directly if positive else it will return 0 removing unnecessary features.
    \item Vital Estimate Generation (1 Dense Layer): Final layer that produces the resulting heart rate or breathing rate estimate.
    
\end{itemize}

\begin{figure*}[!t]
    \centering
    \includegraphics[width=1\linewidth]{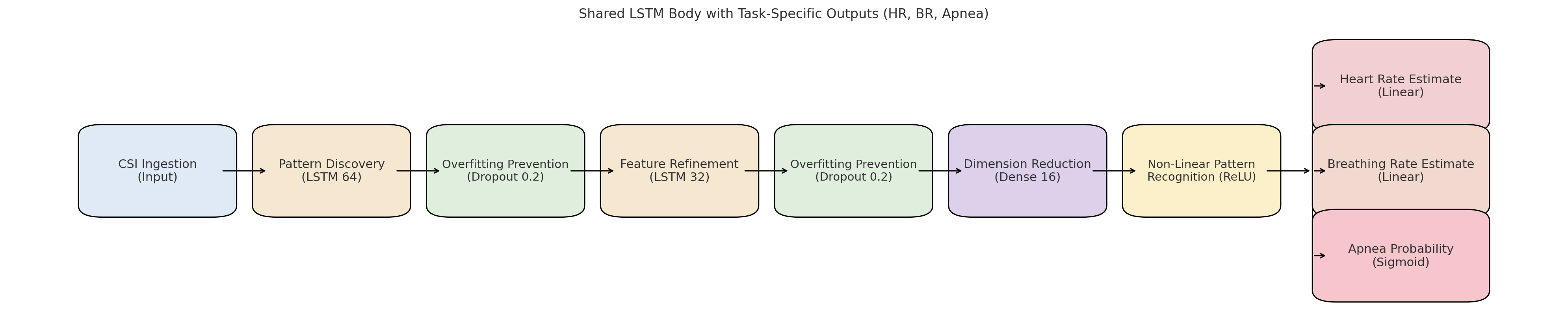}
    \caption{PulseFi LSTM Architecture}
    \label{fig:Lstm_arch}
\end{figure*}

\subsubsection{Model Training}
\label{sec:model_training}
We train our model using the Adaptive Moment Estimation optimizer (ADAM)~\cite{kingma2014adam}, as it is proven to do well with large datasets and high dimensional parameter spaces. ADAM is a widely adopted deep learning approach that uses an optimization algorithm combining momentum and adaptive learning rates to efficiently update the LSTM's network weights. In our current implementation for both heart and breathing rates, the learning rate was empirically set to 0.001 as it provided a balanced trade-off between convergence speed and stability. For our loss function, we use Mean Squared Error (MSE) as described in Equation~\ref{eq:MSE}, which penalizes larger errors more heavily:

\begin{equation}
\mathrm{MSE} = \frac{1}{n} \sum_{i=1}^{n} (y_i - \hat{y}_i)^2
\label{eq:MSE}
\end{equation}

To avoid overfitting, we use a technique called early stopping which works as follows. We check how well the model performs on a validation dataset created by taking 20\% of the training set. At every training round (epoch), we check if the performance stops improving and if that does not improve for 10 consecutive epochs, we stop training early, preventing the model from learning noise instead of useful patterns. Additionally, we use a learning rate reduction strategy that halves the learning rate if there is no improvement in validation loss for 5 consecutive epochs. 

\section{Apnea Detection}
The detection of apnea events, which are characterized by temporary pauses in breathing, uses an architectural foundation similar to the heart and breathing rate monitoring module. It also employs an LSTM to process CSI data and decide whether an apnea event has occurred. We could not use a moving average on top of the pre-existing breathing rate model because the E-Health dataset did not offer breathing rate ground truth. The primary difference between the apnea detection model compared to the heart rate and breathing rate architecture occurs in the final layers and the training objectives as follows:
\begin{itemize}
\item Non Linear Pattern Recognition and Apnea Probability: Instead of producing a continuous vital rate estimate, the final Dense layer is configured for binary classification. It consists of a single neuron whose output is passed through a Sigmoid activation function. The Sigmoid function maps the output to a range between 0 and 1, representing the predicted probability of an apnea event occurring within the input window. This replaces the non linear pattern recognition used in the final layer for regression.
\item Loss Function: The model is trained using Binary Cross-Entropy (BCE) loss. BCE measures the dissimilarity between the predicted probability (from the Sigmoid output) and the actual binary label (0 for normal breathing, 1 for apnea event). This is in contrast with the Mean Squared Error (MSE) loss function used for heart or breathing rate estimation. 
\end{itemize}

Additionally, performance evaluation shifts from regression-focused metrics like MAE and MAPE used for heart- and breathing rate estimation to standard classification metrics, such as accuracy, sensitivity, and specificity. Section VII.D describes in detail the metrics we use to evaluate PulseFi's apnea detection.

The model training process for apnea detection is similar to what we used for heart- and breathing rate estimation, including the use of ADAM, early stopping and learning rate reduction, all of which are described in~\ref{sec:model_training}. 

\section{Experimental Methodology}

\subsection{CSI Data Collection}
We evaluated PulseFi using two different datasets\footnote{Consent was obtained for the data collection.}. The first, the ESP-CSI dataset, was collected locally using two ESP32's, each with a single antenna. As the second dataset, we use the EHealth dataset~\cite{galdino2023ehealth_csi} which was collected by researchers in Brazil using a Raspberry Pi with a single antenna. 
We describe these datasets in more detail below.

\subsubsection{ESP-CSI Dataset}
We collected the ESP-CSI dataset from seven participants (5 men, 2 women) in a semi-controlled indoor environment. Data were collected within a room of a public library. The data were collected using two ESP32 devices, one as the transmitter and the other as the receiver at a sampling rate of 80Hz, 20MHz bandwidth with 64 subcarriers. The devices were placed on custom 3D-printed stands and repositioned to collect data at different distances. The participants sat in a chair between the devices and wore a pulse oximeter on their finger to collect ground truth information. Fig 6 shows our data collection setup. The data from this dataset was used for heart rate estimation and breathing rate estimation, not apnea.

\subsubsection{E-Health Dataset}
The E-Health dataset~\cite{galdino2023ehealth_csi} contains CSI collected from 118 participants (88 men, 30 women) in a controlled indoor environment. The participants' age ranged from 18 to 64 years (mean age 22.38 years), height from 152 to 198 cm, and weight from 40 to 116 kg. The collection was carried out in a 3m x 4m room.
The data collection setup was configured with the following equipment:
\begin{itemize}
\item 5GHz (channel 36) Wi-Fi router at 80MHz bandwidth serving as transmitter.
\item Laptop serving as receiver.
\item Raspberry Pi 4B (single antenna) with NEXMON firmware for CSI data collection (234 subcarriers).
\item Samsung Galaxy Watch 4 worn by participants for ground truth heart rate data.
\end{itemize}

The devices were spaced one meter apart, with the router and participant on opposite sides and the Raspberry Pi equidistant from both, as shown in Fig 6.

\begin{figure}[htbp]
    \centering
    \includegraphics[width=0.4\linewidth]{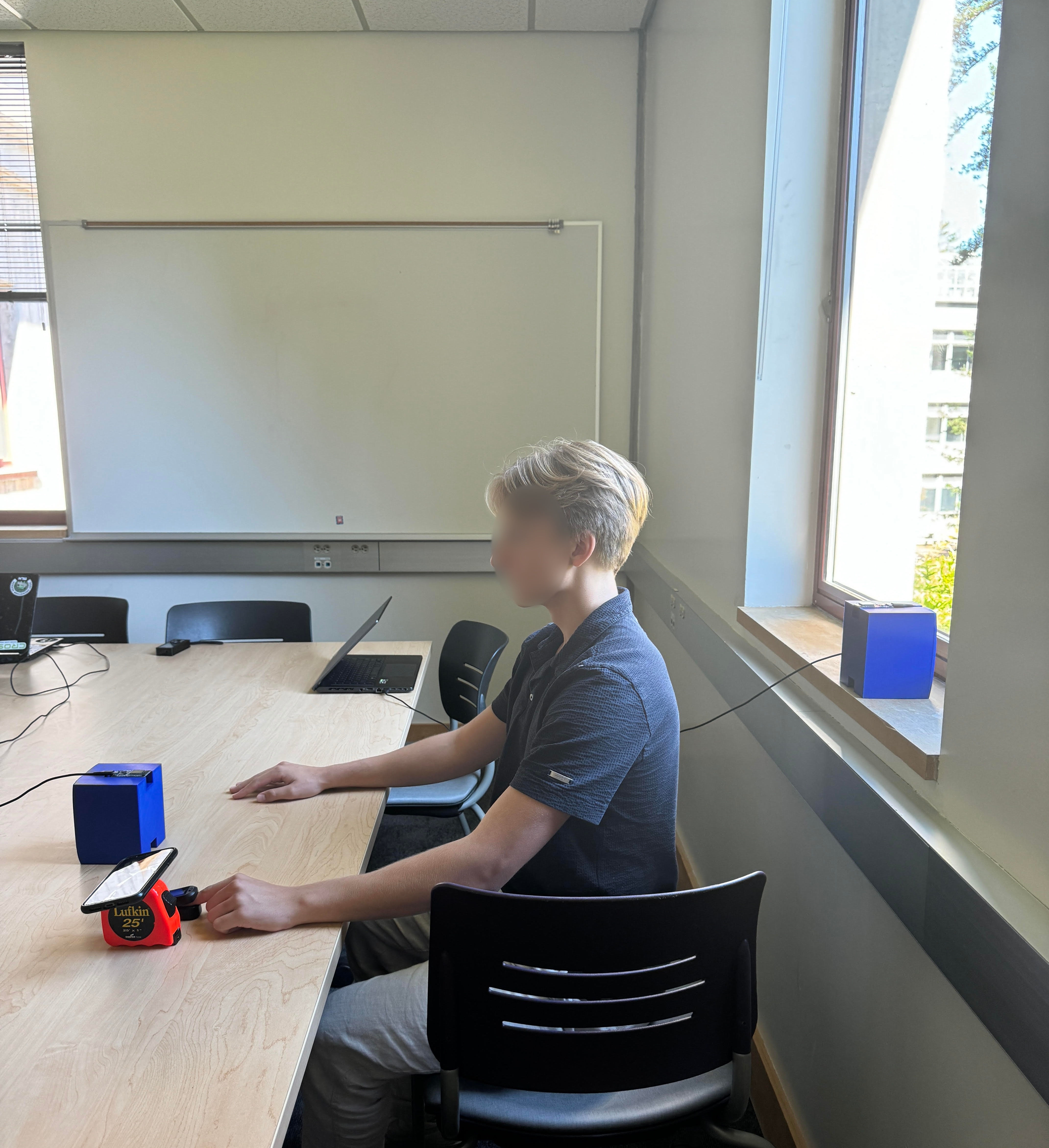}
    \includegraphics[width=0.36\linewidth]{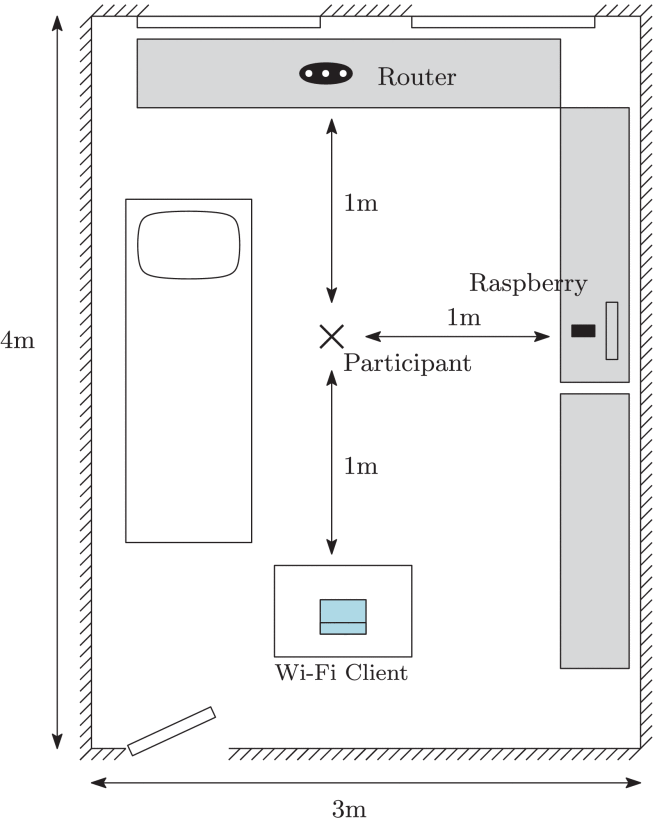}
    \caption{ESP-CSI (left) and eHealth (right) Dataset Collection Setup}
    \label{fig:custom_data_colec}
\end{figure}

Each participant performed 17 standardized positions/activities, with each position held for 60 seconds. Fig 7 illustrates all the EHealth participants' positions and activities. Positions 5, 7, 9, 11, and 13 use what is described as alternate breathing (normal breathing for 20s, hold breath for 10s) and these positions act as the positive samples for apnea detection. Due to extraction issues reported by dataset creators, only heart rate labels are present for the eHealth dataset, not breathing rate. Therefore, the eHealth dataset was used only for heart rate monitoring, and apnea detection. The Wi-Fi client transmitted pings to the router at 136ms intervals (sampling rate of around 7.4), chosen specifically to allow for vital sign monitoring (breathing rate and heart rate is typically observed between 0.1Hz and 3.5Hz).

\subsection{LSTM Model Training and Validation}
PulseFi's LSTM models were trained and validated following standard, widely-used procedures where, for both datasets, 64\% of the data is used for training, 16\% for validation, and the remaining 20\% for testing. The model for each window size was trained and tested three times and the averages are reported. Each time a model is trained, the data is shuffled. This follows standard practice~\cite{ma2020dual_refinement} to allow the model to find a hyper-parameter configuration that generalizes across diverse datasets.

\subsection{Vital Sign Estimation Evaluation Metrics}

To assess PulseFi's performance to monitor and estimate vital signs, we measure a few key metrics. We use Mean Absolute Error (MAE) which is calculated as the average error of the predictions (see Equation~\ref{eq:mae}) and Mean Absolute Percentage Error (MAPE) which expresses accuracy as a percentage of the error (see Equation~\ref{eq:mape}). MAPE provides a measure of prediction accuracy that is scale-independent, allowing for easier interpretation across different ranges of heart rates. We use both metrics as some existing approaches report MAE as their performance parameter, while others use MAPE. We also calculate the percentage of estimations under 1.5 beats per minute error and the percentage under 0.75 breaths per minute, which are standard clinical thresholds. 

\begin{equation}
    \text{MAE} = \frac{1}{n}\sum_{i=1}^{n}|y_i - \hat{y}_i|
    \label{eq:mae}
\end{equation}

\begin{equation}
    \text{MAPE} = \frac{100\%}{n}\sum_{i=1}^{n}\left|\frac{y_i - \hat{y}_i}{y_i}\right|
    \label{eq:mape}
\end{equation}

\begin{equation}
    \text{Accuracy}_{1.5} = \frac{100\%}{n}\sum_{i=1}^{n}I(|y_i - \hat{y}_i| \leq 1.5)
    \label{eq:accuracy_hr}
\end{equation}

\begin{equation}
    \text{Accuracy}_{0.75} = \frac{100\%}{n}\sum_{i=1}^{n}I(|y_i - \hat{y}_i| \leq 0.75)
    \label{eq:accuracy_br}
\end{equation}

\subsection{Apnea Detection Evaluation Metrics}\label{AA}
The metrics chosen to evaluate PulseFi's apnea detection are described below. They are consistent with prior approaches, along with general medical standards. 
 \begin{itemize}
    \item Accuracy: Percentage of correctly classified instances (both apnea and normal breathing).
    \item Sensitivity: Percentage of actual apnea events that were correctly identified by the model. High sensitivity is crucial for minimizing missed detections.
    \item Specificity: Percentage of actual normal breathing periods that were correctly identified. High specificity minimizes false positives.
    \item Cohen's Kappa Coefficient: The agreement between the predicted and actual values, while correcting for agreement that might occur purely by chance. It provides a more robust measure than simple accuracy when class distributions are imbalanced.

\end{itemize}
 
\section{Results}
\subsection{Heart Rate}

\subsubsection{ESP-HR-CSI Data Performance}
\begin{figure}[htbp]
    \centering
    \includegraphics[width=\columnwidth]{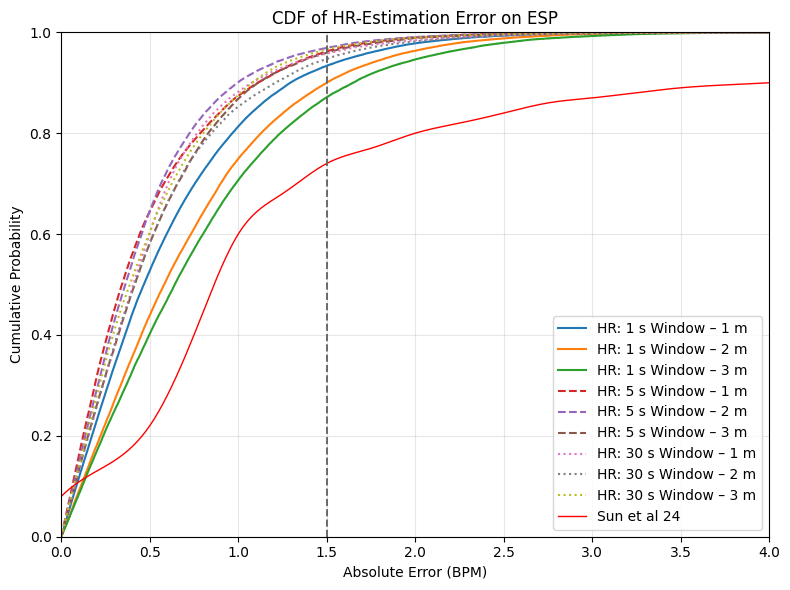}
    \caption{CDF of MAE on ESP-HR-CSI data for different length Windows}
    \label{fig:hr_cdf}
\end{figure}

Table \ref{table:combined_metrics} and Fig \ref{fig:hr_cdf} show PulseFi's heart rate estimation performance across different window sizes and transmitter-receiver distances. The system achieves its best performance at 5-second windows with an MAE of 0.50 BPM and MAPE of 0.51\%. Performance significantly improves from 1-second windows (MAE: 0.90) to 5-second windows (MAE: 0.50), then plateaus at longer durations. This improvement occurs because longer windows capture a larger number of heartbeats, providing a richer temporal context for the pulse waveform. At typical heart rates of 60-90 BPM, a 1-second window may contain zero or only one heartbeat, making reliable estimation impossible. A 5-second window guarantees multiple cardiac cycles (5-7 peaks), providing sufficient data for the LSTM to distinguish true heartbeat patterns from noise. The plateau beyond 5 seconds suggests we've reached the accuracy limits caused by the ESP32's 64-subcarrier resolution and ground truth measurement precision from the pulse oximeter. PulseFi maintains consistent accuracy across varying distances, with only 0.05 MAE variation between 1m and 3m separations using 5-second windows (Fig \ref{fig:hr_cdf}). This stability has not been found with existing peak-detection approaches. Sun \cite{sun2024noncontact_hr} reported MAE doubling from 0.80 to 1.75 BPM between 1m and 3m, while Khamis \cite{khamis2018cardiofi} showed degradation from 1.14 to nearly 3.0 BPM. The LSTM architecture handles signal attenuation and increased noise at greater distances more robustly than traditional frequency-domain peak detection, which fails when cardiac signals weaken below the noise floor. Ten-fold cross-validation with a 15s window gives MAE of 0.892 (std: 0.40), with higher variability reflecting the diverse heart rate ranges across our seven participants. When using 5-second windows or longer, 97.95\% of estimates fall within the 1.5 BPM clinical accuracy threshold. Table II compares PulseFi's performance against state-of-the-art methods, showing state-of-the-art accuracy while using more accessible single-antenna hardware.

\begin{figure}[htbp]
    \centering
    \includegraphics[width=1\columnwidth]{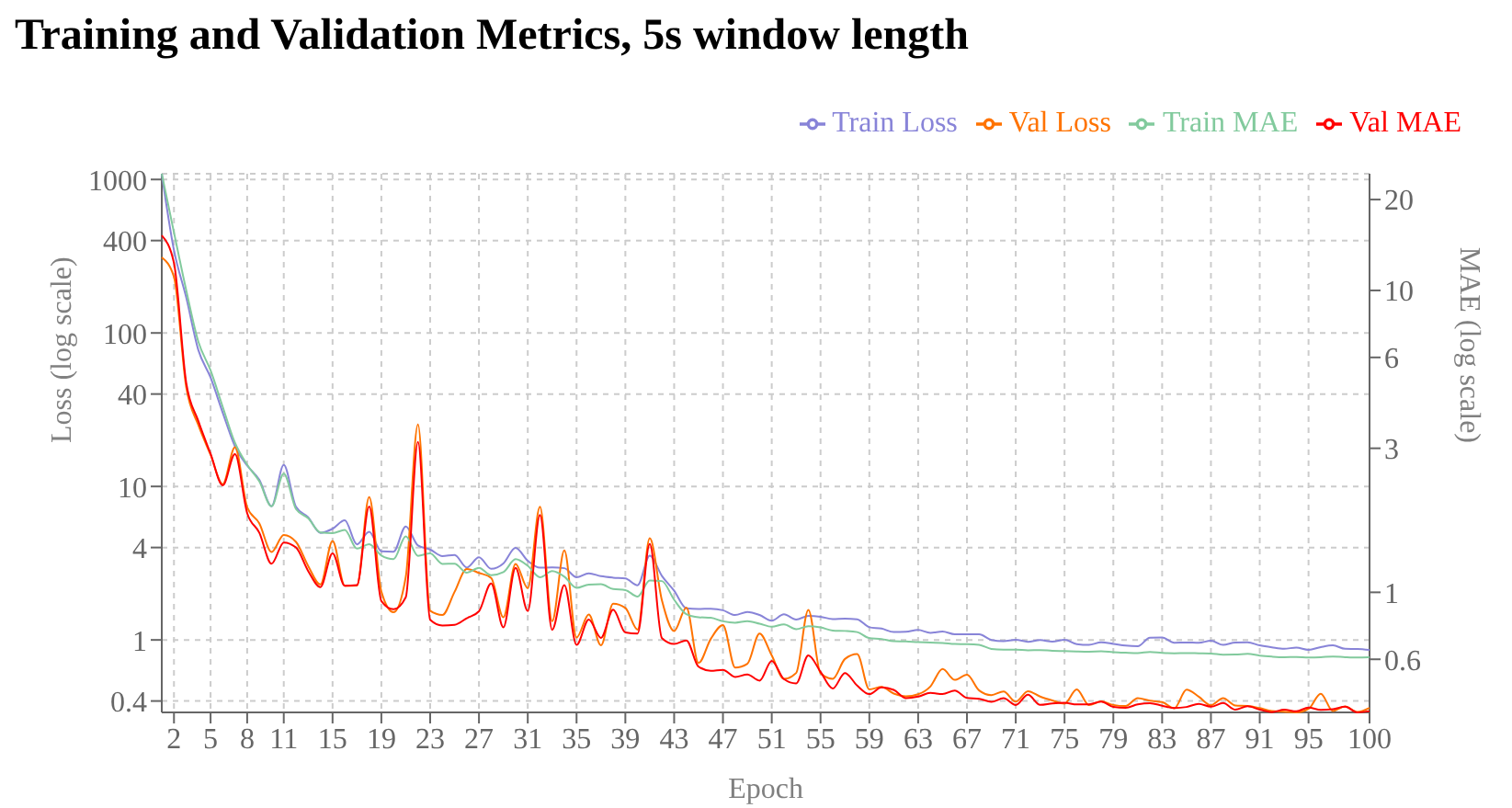}
    \caption{5s Window Length training/validation log on ESP-HR-CSI}
    \label{fig:hr_training_log}
\end{figure}

\begin{table*}[htbp]
\caption{Performance on ESP-HR-CSI and EHealth Data for Heart Rate estimation}
\label{table:combined_metrics}
\begin{center}
\small
\setlength{\tabcolsep}{3.5pt}
\begin{tabular}{|l|c|c|c|c|c|c|}
\hline
\multirow{2}{*}{Win (s)} & \multicolumn{3}{c|}{ESP-HR-CSI} & \multicolumn{3}{c|}{EHealth} \\
\cline{2-7}
& MAE & MAPE & $<$1.5(\%) & MAE & MAPE & $<$1.5(\%) \\
\hline
1 & 0.90$\pm$0.30 & 0.85$\pm$0.29 & 90.75$\pm$9.49 & 6.11$\pm$0.05 & 7.36$\pm$0.07 & 16.22$\pm$0.12 \\
\hline
5 & \textbf{0.50$\pm$0.12 }& \textbf{ 0.51$\pm$0.14 }& \textbf{97.95$\pm$3.81} & 0.76$\pm$0.05 & 0.91$\pm$0.06 & 88.46$\pm$1.78 \\
\hline
10 & 0.51$\pm$0.01 & 0.63$\pm$0.03 & 95.55$\pm$1.35 & 0.27$\pm$0.03 & 0.33$\pm$0.03 & 99.38$\pm$0.10 \\
\hline
15 & 0.51$\pm$0.10 & 0.54$\pm$0.15 & 97.28$\pm$2.78 & 0.26$\pm$0.04 & 0.31$\pm$0.05 & 99.38$\pm$0.18 \\
\hline
20 & 0.56$\pm$0.09 & 0.57$\pm$0.10 & 96.71$\pm$1.98 & 0.22$\pm$0.04 & 0.27$\pm$0.05 & 99.53$\pm$0.11 \\
\hline
25 & 0.58$\pm$0.13 & 0.54$\pm$0.15 & 97.13$\pm$3.80 & 0.18$\pm$0.04 & 0.22$\pm$0.05 & \textbf{99.68$\pm$0.11} \\
\hline
30 & 0.55$\pm$0.05 & 0.66$\pm$0.04 & 95.60$\pm$1.42 & \textbf{0.17$\pm$0.01} & \textbf{0.21$\pm$0.02} & 99.65$\pm$0.03 \\
\hline
\end{tabular}
\end{center}
\end{table*}

\subsubsection{EHealth Data Performance}
The EHealth dataset evaluation shows how window size and participant positioning affect accuracy. PulseFi achieves best performance at 30-second windows with MAE of 0.17 BPM and MAPE of 0.21\%. Unlike the ESP-HR-CSI results, performance continues improving through 30-second windows (Table \ref{tab:apnea_comparison_styled}). At 5s: MAE 0.76, at 10s: MAE 0.27, at 30s: MAE 0.17. This extended improvement stems from two factors: (1) the Raspberry Pi's 234 subcarriers provide higher resolution CSI data, and (2) the larger, more diverse dataset enables the LSTM to learn robust long-term dependencies. The 10-second window achieves 99.38\% of estimates within 1.5 BPM, rising to 99.65\% at 30 seconds. Table \ref{table:hr_table_mae} shows PulseFi maintains consistent performance across all 17 positions (MAE range: 0.21-0.25 BPM). This represents a 93.8\% improvement over the previous approach on this dataset\cite{gouveia2024parameter_tuning}, which showed MAE variations up to 6.67 BPM across positions. 

Ten-fold cross-validation (15s window) achieved MAE of 0.283 (std: 0.053), demonstrating model generalization across the large, diverse participant pool. Compared to existing work (Table \ref{table:comparison}), PulseFi shows 71.7\% improvement over the previous machine learning approach \cite{liu2022human_jbhi} and 78.75\% improvement over peak-detection methods \cite{sun2024noncontact_hr}, while requiring only single-antenna hardware.

\begin{table*}[!htbp]
\caption{Comparison of Different Heart Rate Estimation Methods}
\label{table:comparison}
\centering
\begin{tabular}{|l|c|c|c|c|c|c|c|}
\hline
\multirow{2}{*}{Name} & \multicolumn{5}{c|}{Parameters} & \multirow{2}{*}{MAE} & \multirow{2}{*}{MAPE (\%)} \\
\cline{2-6}
& Method & Window (s) & Chip & Ant. & Part. & & \\
\hline
Wang \cite{wang2020csi} & Peaks & Sliding & Intel 5300 & 3 & 4 & 1.19 & 95.5 \\
Sun \cite{sun2024noncontact_hr} & Peaks & 12-20 & Intel 5300 & 3 & 9 & 0.80 & 97.1 \\
Zhang \cite{zhang2023wital} & Peaks & 30 & Intel 5300 & 3 & 1 & 3.53 & 94.7 \\
Khamis \cite{khamis2018cardiofi} & Peaks & 20 & Intel 5300 & 3 & 4 & 1.14 & NA \\
Gouveia \cite{gouveia2024parameter_tuning} & Peaks & Case specific & Raspberry Pi & 1 & 59 & 2.72 & NA \\
Liu \cite{liu2022human_jbhi} & CNN & 50 & TL-WDR 4300 & 2 & 1 & 0.60 & 98.58 \\
Tsubota \cite{tsubota2021biometric} & Peaks & NA & Intel 5300 & 3 & NA & 1.00 & 98.5 \\
\textbf{PulseFi (ESP-HR-CSI)} & \textbf{LSTM} & \textbf{5} & \textbf{ESP32} & \textbf{1} & \textbf{7} & \textbf{0.50} & \textbf{99.49} \\
\textbf{PulseFi (EHealth)} & \textbf{LSTM} & \textbf{10} & \textbf{Raspberry Pi} & \textbf{1} & \textbf{118} & \textbf{0.27} & \textbf{99.67} \\
\hline
\end{tabular}
\end{table*}

\begin{table}[!t]
\centering
\caption{20s window performance for HR per position compared to State of The Art (SOTA) on EHealth data}
\label{table:hr_table_mae}
\begin{tabular}{|c|c|c|}
\hline
\textbf{Position} & \textbf{MAE$_{\text{SOTA}}$} & \textbf{MAE$_{\text{proposed}}$} \\
\hline
1 & 1.44 & 0.22 \\
\hline
2 & 0.29 & 0.22 \\
\hline
3 & 0.73 & 0.23 \\
\hline
4 & 1.21 & 0.22 \\
\hline
5 & 2.60 & 0.22 \\
\hline
6 & 4.29 & 0.22 \\
\hline
7 & 6.02 & 0.22 \\
\hline
8 & 6.37 & 0.22 \\
\hline
9 & 6.74 & 0.22 \\
\hline
10 & 0.07 & 0.21 \\
\hline
11 & 0.30 & 0.22 \\
\hline
12 & 1.24 & 0.21 \\
\hline
13 & 2.86 & 0.21 \\
\hline
14 & 2.04 & 0.22 \\
\hline
15 & 3.96 & 0.23 \\
\hline
16 & 1.79 & 0.25 \\
\hline
17 & 4.60 & 0.24 \\
\hline
Average & 2.72 & 0.22\\
\hline
\end{tabular}
\end{table}

\subsection{Breathing Rate}
Table \ref{table:esp_br_csi_metrics} and Figure \ref{fig:br_cdf} show breathing rate estimation results on the ESP-BR-CSI dataset. PulseFi achieves best performance at 20-second windows with MAE of 0.09 breaths/min and MAPE of 0.59\%. There was no statistically significant difference in MAE or MAPE after this point. Breathing rate shows even greater sensitivity to window size than heart rate. Performance improves steadily from 1s (MAE: 0.61) through 20s (MAE: 0.09), reflecting the slower respiratory cycle. Typical breathing rates of 12-20 breaths/min mean 3-5 second periods between breaths. A 1-second window rarely captures even a single complete breath, while a 20-second window contains 4-6 full respiratory cycles, allowing for more robust estimation and reducing the impact of irregular breathing patterns. The continued improvement beyond heart rate's plateau point happens because respiratory signals, though stronger in amplitude, have lower frequency components that require longer windows for accurate estimation. Breathing rate estimation maintains consistency across distances, with only 0.01 MAE variation between 1m and 3m using 20-second windows. The stronger chest movement signal during respiration remains above the noise floor even at 3m separation, allowing the LSTM to reliably extract breathing patterns. Using windows $\geq$5 seconds, 97.8\% of estimates fall within the 0.75 breaths/min clinical threshold, rising to 99.6\% at 20 seconds. Ten-fold cross-validation (15s window) gives a MAE of 0.103 (std: 0.009), showing more stable performance than heart rate due to breathing's narrower range. Table \ref{table:breathing_rate_comparison} compares PulseFi against existing approaches, demonstrating a 10\% improvement over the state-of-the-art best result~\cite{guo2023breatheband} while using lower cost, more accessible hardware.

\begin{table}[htbp]
\caption{Performance on ESP-BR-CSI Data for Breathing Rate estimation}
\label{table:esp_br_csi_metrics}
\begin{center}
\small
\setlength{\tabcolsep}{3.5pt}
\begin{tabular}{|l|c|c|c|}
\hline
\multirow{2}{*}{Win (s)} & \multicolumn{3}{c|}{ESP-BR-CSI} \\
\cline{2-4}
& MAE & MAPE & $<$0.75(\%) \\
\hline
1 & 0.61$\pm$0.04 & 3.85$\pm$0.25 & 73.5$\pm$2.9 \\
\hline
5 & 0.20$\pm$0.04 & 1.24$\pm$0.28 & 97.8$\pm$1.0 \\
\hline
10 & 0.12$\pm$0.01 & 0.76$\pm$0.09 & 99.5$\pm$0.2 \\
\hline
15 & 0.11$\pm$0.01 & 0.71$\pm$0.06 & 99.5$\pm$0.1 \\
\hline
20 & \textbf{0.09$\pm$0.00} & \textbf {0.59$\pm$0.01} & \textbf{99.6$\pm$0.0} \\
\hline
25 & 0.10$\pm$0.02 & 0.63$\pm$0.13 & 99.6$\pm$0.2 \\
\hline
30 & 0.09$\pm$0.01 & 0.60$\pm$0.04 & 99.6$\pm$0.0 \\
\hline
\end{tabular}
\end{center}
\end{table}

\begin{figure}[htbp]
    \centering
    \includegraphics[width=1\columnwidth]{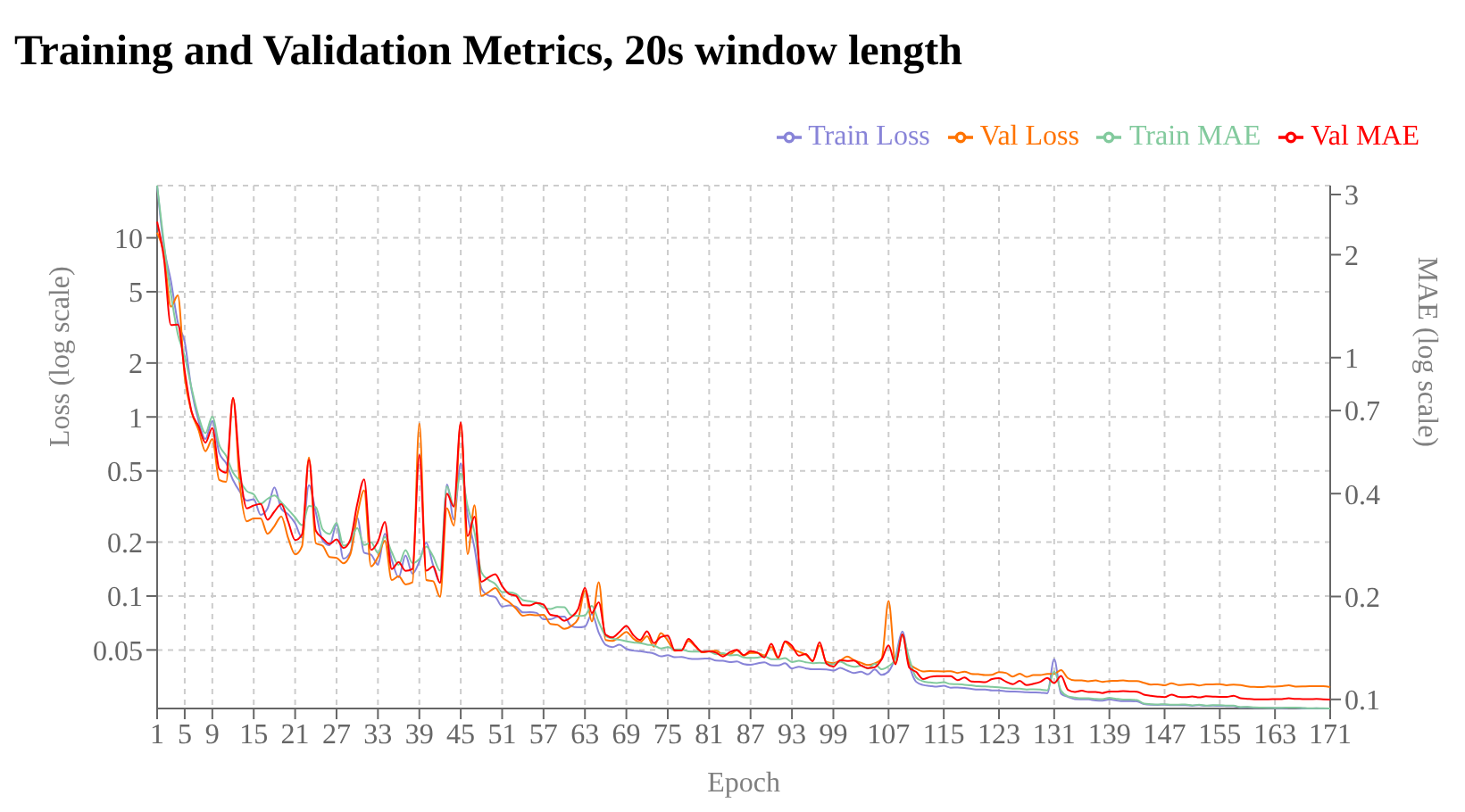}
    \caption{20s Window Length training/validation log on ESP-BR-CSI}
    \label{fig:breath_training_log}
\end{figure}

\begin{figure}
    \centering
    \includegraphics[width=\columnwidth]{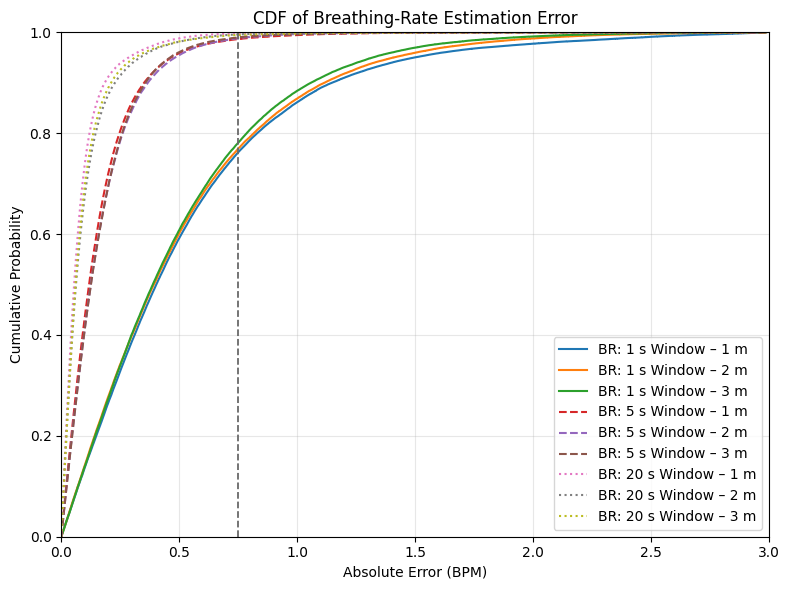}
    \caption{CDF of MAE on ESP-BR-CSI data for different distances and window lengths}
    \label{fig:br_cdf}
\end{figure}

\begin{table*}[htbp]
\caption{Comparison of Different Breathing Rate Estimation Methods}
\label{table:breathing_rate_comparison}
\begin{center}
\small
\setlength{\tabcolsep}{3.5pt}
\begin{tabular}{|l|c|c|c|c|c|c|}
\hline
\multirow{2}{*}{Name} & \multicolumn{5}{c|}{Parameters} & \multirow{2}{*}{MAE} \\
\cline{2-6}
& Method & Window (s) & Chip & Antenna & Participant & \\
\hline
Li~\cite{li2021complexbeat} & Peaks & 30 & AR9462 & 2 & 20 & 0.25 \\
\hline  
Wang~\cite{wang2020resilient} & Peaks & 30 & Intel 5300 & 3 & 1 & 0.25 \\
\hline
Wang~\cite{wang2017tensorbeat} & Peaks & 30 & Intel 5300 & 3 & 5 & 0.25 \\
\hline
Duo~\cite{dou2021full} & Peaks & NA & Intel 5300 & 3 & 1 & 0.14 \\
\hline
Liu~\cite{liu2022human_jbhi} & CNN & 50 & TL-WDR 4300 & 2 & 1 & 0.20 \\
\hline
Guo~\cite{guo2023breatheband} & HHM & 30 & Unknown & 3 & 1 & 0.10 \\
\hline
Wang~\cite{wang2020csi} & Peaks & 3-5 & Intel 5300 & 3 & 4 & 0.25 \\
\hline
\textbf{PulseFi} & \textbf{LSTM} & 20 & \textbf{ ESP32} & \textbf{1} & 7 & \textbf{0.09} \\
\hline
\end{tabular}
\end{center}
\end{table*}

\subsection{Apnea Detection}
Table~\ref{tab:apnea_metrics} shows apnea detection performance across different window sizes. Using a 10-second window, PulseFi achieves 99.4\% accuracy, 95.7\% sensitivity, 99.7\% specificity, and 95.9 Kappa coefficient. PulseFi's near-perfect classification comes from apnea's distinctive signal characteristics. Complete stop of respiratory movement creates a clear absence of the cyclical CSI changes found during normal breathing. Unlike heart and breathing rate estimation which require precise frequency measurement, apnea detection is a binary classification task where signal presence/absence provides a strong discriminative feature. Unlike vital rate estimation, apnea detection shows minimal performance variation for different window sizes (Table~\ref{tab:apnea_metrics}). Comparing the 10s to 5s window, there is a modest improvement of 0.3\% in accuracy and 1.2 Kappa suggesting even short observation periods are sufficient for reliable apnea detection. Looking across positions, highest accuracy was positions 1 and 2 at 99.7\% and lowest was position 16 with 99.2\% when using a 10s window, showing that detection accuracy is relatively immune to position, similar to what was seen with heart rate. Table~\ref{tab:apnea_comparison_styled} shows PulseFi's apnea detection performance compared to existing wireless signal based approaches, where we demonstrate a 10\% Kappa improvement over the previous best Wi-Fi CSI approach~\cite{liu2014wisleep} and better performance than camera~\cite{abad2016video_apnea} and radar~\cite{beattie2013accurate_scoring} systems.

\begin{table}[htbp]
\centering
\caption{Apnea Detection Performance Per Window Sizes}
\begin{tabular}{|c|c|c|c|c|}
\hline
\textbf{Time} & \textbf{Accuracy} & \textbf{Sensitivity} & \textbf{Specificity} & \textbf{Kappa} \\
\hline
5s & 99.1$\pm$0.2 & 95.3$\pm$0.1 & 99.6$\pm$0.1 & 94.7$\pm$0.2 \\
10s & \textbf{99.4$\pm$0.1} & \textbf{95.7$\pm$0.1} & \textbf{99.7$\pm$0.0} & \textbf{95.9$\pm$0.1} \\
\hline
\end{tabular}
\label{tab:apnea_metrics}
\end{table}

\begin{table*}[!t]
  \caption{Comparison of Sleep Apnea Detection Methods using Non-Contact Sensors}
  \label{tab:apnea_comparison_styled} 
  \centering
  \footnotesize 
  \begin{tabular}{|l|l|l|c|c|c|c|}
    \hline
    Reference & Measurement Mechanism & Approach & Accuracy (\%) & Sensitivity (\%) & Specificity (\%) & Kappa \\
    \hline
    Abad et al. ~\cite{abad2016video_apnea} & Video Camera (SleepWise) & Image Processing & 98 & 100 & 83 & 0.79 \\ \hline
    Kang et al. \cite{kang2020noncontact_apnea} & IR-UWB Radar & Statistical Thresholding & 93 & 100 & 92 & -- \\ \hline
    Choi et al. \cite{choi2022automated_apnea} & 60GHz FMCW Radar & CRNN & 75 & 63 & -- & 0.72 \\ \hline
    Zhuang et al. \cite{zhuang2022accurate_apnea} & 24GHz FMCW Radar & Random Forest + Signal Proc. & 95.5 & 72.6 & 97.3 & 0.68 \\ \hline
    Koda et al. \cite{koda2021radar_apnea} & 24GHz FMCW Radar & SVM & 66.1 & 22.3 & 69.6 & -- \\ \hline
    Wang et al. \cite{liu2014wisleep} & Wi-Fi CSI (WiSleep) & Machine Learning (SVM) & 89.5 & 92.6 & 88.9 & -- \\ \hline
    Beattie et al. \cite{beattie2013accurate_scoring} & Doppler Radar & Signal Processing & 95.1 & 92.7 & 98.4 & 0.86 \\ \hline
    \textbf{PulseFi} & \textbf{Wi-Fi CSI} & \textbf{Signal Proc + LSTM}  & \textbf{99.4}  & \textbf{95.7} & \textbf{99.7}  &  \textbf{95.9} \\ \hline
  \end{tabular}
\end{table*}

\subsection{Computational Efficiency}
PulseFi's compact architecture allows deployment on resource-constrained devices. The complete LSTM models contain 46,113 trainable parameters and require 500-600KB of storage. Table \ref{tab:cpu_deployment_metrics} shows CPU inference performance metrics for heart rate models across different window sizes. With a 5-second window using the ESP32 dataset, inference requires approximately 36 million floating-point operations. The ESP32 running at 240MHz can complete this computation in ~0.15 seconds, allowing real-time monitoring with updates every 5 seconds using a very low-cost device. On Raspberry Pi hardware (similar computational capacity to Google Colab CPU in our tests), the 5-second window model achieves 2,335 predictions/second throughput with 26ms batch mean latency. Even the 30-second window model, despite 2,400-sample sequences, maintains a 299 predictions/second throughput.

\begin{table*}
\centering
\caption{CPU Deployment Performance Metrics for Heart Rate Models}
\label{tab:cpu_deployment_metrics}
\begin{tabular}{@{}lccccccc@{}}
\toprule
Model & Seq & Batch & Cold Start & Total Time & N & Throughput & Batch Mean \\
& Len & Size & (s) & (s) & Preds & (preds/s) & (ms) \\
\midrule
HR: 1s Window & 80 & 64 & 0.170 & 36.65 & 337,501 & 9,209.72 & 6.36 \\
HR: 5s Window & 400 & 64 & 0.044 & 144.36 & 337,181 & 2,335.70 & 26.09 \\
HR: 30s Window & 2,400 & 64 & 0.172 & 1,119.36 & 335,181 & 299.44 & 207.17 \\
\bottomrule
\end{tabular}
\end{table*}

\section{Discussion}

PulseFi demonstrates that accurate cardiopulmonary monitoring is achievable using commodity single-antenna hardware through LSTM learning. Our results show that 97.95\% of heart rate estimates and 99.6\% of breathing rate estimates fall within established clinical accuracy thresholds (±1.5 BPM and ±0.75 breaths/min respectively), meeting medical standards that typically accept 1-5\% error margins for vital sign measurements. The system's consistent performance across distances (1-3m) and participant positions covers a large gap in prior Wi-Fi sensing work, which generally degraded as distance increased and position changed. Our results show a hardware imposed accuracy limitations with the ESP32. Performance plateaus at 5 second window (MAE 0.5BPM) regardless of longer window size, suggesting we've reached the maximum accuracy achievable with only 64 subcarriers. In contrast, the Raspberry Pi's better performance (MAE 0.17 at 30 second window) shows how higher subcarrier count directly improves accuracy. The Raspberry Pi's 234 subcarrier gives 3.6x higher frequency resolution than the ESP32's 64 subcarriers, allowing for more detailed capture of subtle cardiac variations. Importantly, this performance gain required no modifications to the PulseFi algorithm suggesting that future implementations using IEEE 802.11bf standard should naturally achieve better accuracy without requring algorithm redesign.

\section{Conclusion}
In this paper, We presented PulseFi, a novel low-cost system that uses Channel State Information (CSI) to continuously and non-intrusively monitor heart rate, breathing rate and apnea. PulseFi highlights that adequate accuracy can be achieved using low-cost, off-the-shelf commodity hardware amplitude information. Our experimental results using two distinct datasets demonstrate that PulseFi's CSI processing pipeline combined with its custom low-compute Long Short Term Memory (LSTM) neural network is able to monitor heart rate accurately. We also show that PulseFi yields heart monitoring accuracy either comparable or higher when compared with existing systems that employ more specialised hardware and/or require higher computational power. 


\bibliographystyle{IEEEtran}
\bibliography{references}

\end{document}